# Topological Degeneracy Induced by Twisting

Han Peng[1, *], Qiang Wang[1, *, †], Meng Xiao[2, 3, †], Xiayi Wang[1], Shining Zhu[1] and Hui Liu[1, †]

[1]*National Laboratory of Solid State Microstructures, Jiangsu Physical Science Research Center, School of Physics, Collaborative Innovation Center of Advanced Microstructures, Nanjing University, Nanjing 210093, China*

[2]*Key Laboratory of Artificial Micro- and Nano-structures of Ministry of Education and School of Physics and Technology, Wuhan University, 430072 Wuhan, China*

[3]*Wuhan Institute of Quantum Technology, Wuhan 430206, China*

*These authors contributed equally to this work.

†Correspondence address: q.wang@nju.edu.cn; phmxiao@whu.edu.cn; liuhui@nju.edu.cn.

**Abstract**:

**In recent years, twisting has emerged as a new degree of freedom that plays an increasingly important role in Bloch bands of various physical systems. However, there is currently a lack of reports on the non-trivial physics of topological degeneracy in twisted systems. In this work, we investigated the intrinsic physical correlation between twisting and topological degeneracy. We found that twisting not only breaks the symmetry of the system but also introduces topological degeneracy that does not exist under the original symmetric system without twisting. Furthermore, the topological degeneracy can be easily tuned through twisting. This new twist-induced topological degeneracy gives rise to a unique polarization-degenerate birefringent medium, wherein the twist angle acts as a novel degree of freedom for dispersion and polarization management of interface states. Exhibiting fascinating properties and experimental feasibilities, our work points to new possibilities in the research of various topological physics in twisted photonics.**

*Introduction.* –Very recently, the twist angle, as a new degree of freedom, has been widely explored to manipulate quantum materials. The delicate interlayer coupling is controlled by the twist angle, leading to the emergent field of moiré structures[1, 2], e.g., the prominent twisted bilayer graphene[3-7]. The moiré physics has also been extended to classical wave systems[8]. In photonics, twist angle can give rise to exotic phenomenon[9-14], including flat band in bilayer photonic crystals[15-19], phase synchronization in nanolasers[20] and many more[21-24]. To date, most of the works on twist photonics have primarily focused on generating flat bands where interlayer couplings intend to introduce gapped phases. However, to the best of our knowledge, there is currently a lack of reports on the non-trivial physics of topological degeneracy in twisted systems.

Topological degeneracies[25-27](TDs) usually serve as the phase transition points between topological trivial and nontrivial phases, thus play a vital role in identification of various topological phases[28]. Systems exhibiting TDs such as Dirac and Weyl nodes[29, 30] are dubbed as topological semimetals. Typical optical systems confined in one direction can also host TDs[31, 32], provided that certain symmetries are preserved[33, 34]. For example, the Dirac points (DPs) in two-dimensional (2D) photonic honeycomb lattice are protected by time reversal symmetry and inversion symmetry[35] . Once either symmetry is broken, TDs would be lifted, resulting in gapped phases[36, 37]. Therefore, most of the previous works insist on preserving certain symmetries to construct topological degeneracy. There is scarcely any work discussing the opposite physical mechanism about symmetry-breaking induced TDs (in particular, DPs)[38, 39], i.e., TDs emerge when certain symmetry is broken.

In this work, we investigated the intrinsic physical correlation between twisting and topological degeneracy. Our findings indicate that twisting not only breaks the symmetry of the system but also introduces topological degeneracy that is absent in the original symmetric system without twisting. We present a specifically designed system to demonstrate the twist-induced TDs. The structure consists of two anisotropic metasurfaces separated and sandwiched by photonic crystals (PCs). Two anisotropic interfaces states (AISs) are supported at the metasurfaces. They coupled to form twisted bilayer AISs (TBAISs) through the PC in between and a band gap opens. By twisting one of the two anisotropic metasurfaces, the up-down mirror symmetry is broken, and intriguingly, two type-II DPs emerge in the momentum space. In other words, mirror symmetry breaking dictates the presence of the DPs. Meanwhile, the position of the DPs can be shifted by tuning the twist angle. Considering the twist angle as an additional dimension besides the 2D momentum space, the Dirac nodes form two nodal lines. These two nodal lines merge when the twist angle is $\pi/2$, and instead of annihilation, they form a charge-2 Dirac node at the crossing point[38]. We note that, the isofrequency contours at the type-II Dirac nodes are similar to the contours of the uniaxial medium, thus our system support 2D uniaxial interface waves that exhibit birefringence effects. The above results were experimentally verified in samples with different twist angles. The twist angle here can be considered as a new synthetic dimension. Over the past decade, constructing synthetic dimensions as new controllable degree of freedom has gained intense attention across various fields[40]. Our work provides a novel and flexible method to tune TDs through twisting, extending the use of synthetic dimensions as an effective knob for tuning topological semimetal phases. Furthermore, the uniaxial interface waves are applicable in phase matching, mode division and photonic integration.

*Theory of topological degeneracy induced by symmetry breaking.* - As depicted in Fig. 1(a), the TBAIS is constructed from two anisotropic metasurfaces (parallel golden bars), which are sandwiched by three 1D PCs. Each metasurface support one AIS, and two AISs couple with each other through the central PC. The metasurfaces are made of gold nanostripes, whose period ($200nm$) is much smaller than the working wavelength ($>800nm$), and it can be modelled as a homogenous hyperbolic medium[41, 42]. The two metasurfaces can twist relative to each other with an angle $\alpha$, and the coordinate axis are defined as the two diagonal directions (inset in Fig. 1a). The 1D PCs are made of Ta2O5 and SiO2 with thickness $d_A$ and $d_B$ respectively (lower panel of Fig. 1a). The unit cell of the PC in between the metasurfaces is chosen as the $A/2 - B - A/2$ configuration and the number of unit cell is 4, while the two outer PCs is in the $B - A$ configuration. Above design ensures that the AISs only exist between the metasurfaces and the middle PC[43]. The two AISs interact with each other via evanescent waves and form TBAISs. These TBAISs possess mirror symmetry with respect to the central plane only at $\alpha = 0°$ and no mirror symmetry otherwise.

The dispersion of TBAISs is obtained by the transfer matrix method[44]. For AIS at a single metasurface, the dispersion is written as $E = E_0 + ak_x^2 + bk_y^2$, with $a \neq b$ indicating the anisotropy of AISs, $E_0$ being the frequency at $k = 0$. We assume that the eigenfield of the AISs are approximated by that at the Γ point (which is a reasonably good approximation around the Γ point, as shown in Figs. S4 and S5). From this, the angular-dependent coupling between the two AISs is derived by calculating the eigenfields overlap between them, and accordingly, the effective Hamiltonian is[44]:

$$H = \frac{1}{2}\left[E_0 + (a+b)(k_x^2 + k_y^2) + (a-b)(k_x^2 - k_y^2)\cos\alpha\right]\cdot\sigma_0 + (a-b)k_x k_y \sin\alpha \cdot \sigma_3$$
$$+ q\left[\cos\alpha + \frac{1}{2}\left[(k_x^2 + k_y^2)\cos\alpha + k_x^2 - k_y^2\right]\right]\cdot\sigma_1, \#(1)$$

where $\sigma_i (i=0,1,2,3)$ stands for the identity matrix and Pauli matrices. Note here the second term ($\sigma_3$) describes the frequency detuning induced by twist between two uncoupled AISs, while the last term ($\sigma_1$) stands the coupling strength between the two AISs, and $q$ is regarded approximately a constant. Considering the condition of $0° \leq \alpha \leq 90°$, for $k_x = 0$, the Second term vanishes and the coefficient before the last term in Eq. (1) is $q[\cos\alpha + \frac{1}{2}k_y^2(\cos\alpha - 1)]$. There are three cases:

(i) The system is up-down mirror-symmetric, i.e., $\alpha = 0°$, and the term $(\cos\alpha - 1)$ equals to zero. Then the coupling strength remains positive regardless of $k_y$. Consequently, no TD can be found in this case.

(ii) The mirror symmetry is broken by a twist ($0° < \alpha < 90°$), then $\cos\alpha$ is positive and $(\cos\alpha - 1)$ is negative. At $k_y = \pm\sqrt{2\cos\alpha/(1-\cos\alpha)}$, the third term in Eq. (1) equals zero. Therefore, there are two TDs formed at these two points. For $90° < \alpha < 180°$, another two TDs at $(\pm\sqrt{2\cos\alpha/(1-\cos\alpha)}, 0)$ are also found following a similar derivation.

(iii) At $\alpha = 90°$, two TDs merge.

In conclusion, the TDs only emerge when the twist breaks the mirror symmetry in TBAISs.

Figure. 1b plots the dispersion of the TBAISs with $\alpha = 0°$, these two bands are gapped with NO TDs. When the mirror symmetry is broken by a twist (e.g., $\alpha = 70°$), these two bands intersect with each other at $(k_x, k_y) = (0, k_y^{TD})$, as shown in Fig. 1c. A zoom in view of one degenerate point is shown in Fig. 1e. It is clear that the TD tilts in momentum space forming a type-II Dirac point since the tilting parameter is larger than unity[44, 47]. Figures 1f and 1g show the typical horizontal electric field $(E_x, E_y)$ of the two states forming the Dirac point at $\alpha = 70°$.

When further increasing the twist angle, the two type-II Dirac points move towards to the origin of the momentum space, i.e., Γ point. At $\alpha = 90°$ (where the two metasurfaces are perpendicular), these two type-II Dirac points merge into one TD with quadratic dispersions at the Γ point. This TD is classified as a charge-2 Dirac point, which characterized by a $2\pi$ Berry phase when enclosing the DP[44]. Notably, the electric fields are parallel to the nanostripes in a single AIS, therefore these two AISs decouple at the charge-2 Dirac point when $\alpha = 90°$.

*Observation of topological degeneracies with tuned twist angles.* -To experimentally demonstrate the above symmetry-breaking induced TDs, a series of samples with different twist angles is fabricated[44]. It is known that multilayer structure suffers from the inevitable loss, however, the loss term here is approximated as an identity matrix and the DPs remain intact[44]. An SEM image of the cross section for one sample is shown in the lower panel of Fig. 1a. In experiments, we measured the reflection spectrum along different direction across the Γ point, thus mapping the dispersion in the 2D momentum space[48].

Figure 2a shows the sketch of measuring dispersion of the TBAISs at $\alpha = 70°$, where a type-II Dirac point is expected. Four measured reflection spectra at different azimuth angles $\theta$ for this sample are plotted in Fig. 2b. The TBAIS is manifested as reflection dips in our measurement. It is clear that two reflection dip lines intersect with each other at $\theta = 0°$ (i.e., along the $k_y$ axis); while for the other three directions, these two dip lines are always separated, i.e., gapped. Figure 2c shows the corresponding dispersion with $\alpha = 90°$, where a charge-2 Dirac point presents at the Γ point, the dispersions are quadratic, consistent with the effective Hamiltonian. For comparison, theoretical dispersions are shown in Figs. 2b and 2d with gray dashed lines, and the Dirac points are denoted by green dots, which match well with the measured results. The above results confirm the observation of the type-II Dirac point and the charge-2 Dirac point on the TBAISs.

According to the above discussion, these TDs are located on the $k_y$ axis for $0° \leq \alpha \leq 90°$, and on the $k_x$ axis otherwise. Subsequently, the measured reflection spectra along $k_y$ axis at different twist angles $\alpha$ are given in Fig. 3a. We note that the theoretical TDs marked as the green dots move towards to the Γ point as twist angle $\alpha$ increase; and they eventually merge into a charge-2 Dirac point at $\alpha = 90°$. Meanwhile, there is no TD at $\alpha = 0°$ as the two reflection dip lines are separated. Taking the twist angle as a synthetic dimension, these TDs form a nodal chain in the 3D space consisting of $k_x, k_y$ and $\alpha$, with a chain point at $(k_x, k_y, \alpha) = (0, 0, \pi/2)$. For $0 \leq \alpha \leq \pi/2$, with expanded Hamiltonian around the chain point, these TDs are located at

$$k_y = \pm \sqrt{2(\pi/2 - \alpha)}, k_x = 0 \#(2)$$

We collect all the band crossing point in Fig. 3a and replot them in the $(k_x, k_y, \alpha)$ space in Fig. 3b. TDs that found in the experiment are marked as solid magenta dots, which agree well with the theoretically predicted nodal chain (green solid line). Thus, we have experimentally confirmed a chained nodal line in the synthetic space.

*Uniaxial isofrequency contours at topological degeneracies.* -In optics, the isofrequency contours plays a decisive role in various effects, including birefringence and negative refraction. In Figs. 4a-4c, we show the isofrequency contours above, at, and below the TDs of the TBAISs at $\alpha = 75°$. For clarity, the analytical isofrequency contours are plotted with colored lines, where the color denotes the polarization ratio of the corresponding eigenmode. It is observed that the inner (outer) contours are mainly dominated by $E_z(H_z)$ near TDs, leading to a polarization difference between these two contours. We note that the isofrequency contour at the TD (Fig. 4b) is quite similar to those in a uniaxial crystal. Unlike conventional uniaxial medium where the isofrequency contours are intersected by a circle (ordinary waves) and an ellipse (extraordinary waves), here the contours of TBAISs are composed of two elliptical-like contours, indicating that both modes are extraordinary. Such a unique feature implies that the TBAISs can host intriguing birefringent effect when the wave is incident from a structure with isotropic isofrequency contour. To be specific, when light is injected along the optical axis, i.e., TD, the two excited states propagate with the same direction; once deviating from the optical axis, the light beam splits into two with different directions[44]. Furthermore, the uniaxial interface waves of TBAIS can also be tuned by the twist angle, as depicted in Figs. 4d-4f. To the best of our knowledge, this is the first time that tunable birefringent effect has been demonstrated within localized interface waves, which favors applications in integrated mode division, phase matching for interface waves and more.

*Summary and outlook.* -The properties of TBAISs were investigated both theoretically and experimentally. Twist between two metasurfaces breaks the up-down mirror symmetry of the system and leads to TDs. Specifically, type-II and charge-2 Dirac points are observed experimentally. Subsequently, a nodal chain is formed in the $k_x - k_y - \alpha$ synthetic space as confirmed by experiment. Our findings enrich the field of twist photonics[9] and may offer a potential route to demonstrating non-abelian braiding by introducing more layers and twist angles[49-51]. Our work not only demonstrates the new possibility of creating TDs by breaking certain symmetries, but also presents a rather simple and flexible platform for manipulating the polarization and propagation of interface waves.

*Acknowledgments*-This work was financially supported by the National Natural Science Foundation of China grant No. 12334015, 92163216, 92150302, 62288101, 12321161645 and 12274332，the Fundamental Research Funds for the Central Universities (No. 2024300369), the Natural Science Foundation of Jiangsu Province (No. BK20233001) and National Natural Science Fund for Excellent Young Scientists Fund Program (Overseas).

H. L. and M. X. proposed and designed the system. H.P. ran the numerical simulations. H.P. and X.W. carried out the experiment. M.X., Q.W., S.N.Z and H.L. supervised the project. H.P., M.X., Q.W., and H.L. co-wrote the manuscript. All the authors contributed to the analysis and discussion of the results. H. P. and Q. W. contributed equally to this work.

# References

[1] S. Shabani, D. Halbertal, W. J. Wu, M. X. Chen, S. Liu, J. Hone, *et al.* Deep moire potentials in twisted transition metal dichalcogenide bilayers. *Nat. Phys.*, **17,** 720(2021).

[2] C. H. Jin, E. C. Regan, A. M. Yan, M. I. B. Utama, D. Q. Wang, S. H. Zhao, *et al.* Observation of moiré excitons in WSe2/WS2 heterostructure superlattices. *Nature*, **567,** 76(2019).

[3] Y. Cao, V. Fatemi, S. Fang, K. Watanabe, T. Taniguchi, E. Kaxiras, and P. Jarillo-Herrero. Unconventional superconductivity in magic-angle graphene superlattices. *Nature*, **556,** 43(2018).

[4] R. Bistritzer, A. H. Macdonald. Moire bands in twisted double-layer graphene. *Proc. Natl.*

*Acad. Sci. U.S.A.*, **108,** 12233(2011).

[5] Z. D. Song, Z. J. Wang, W. J. Shi, G. Li, C. Fang, and B. A. Bernevig. All Magic Angles in Twisted Bilayer Graphene are Topological. *Phys. Rev. Lett.*, **123,** 036401(2019).

[6] G. Tarnopolsky, A. J. Kruchkov, and A. Vishwanath. Origin of Magic Angles in Twisted Bilayer Graphene. *Phys. Rev. Lett.*, **122,** 106405(2019).

[7] S. Wu, Z. Y. Zhang, K. Watanabe, T. Taniguchi, and E. Y. Andrei. Chern insulators, van Hove singularities and topological flat bands in magic-angle twisted bilayer graphene. *Nat. Mat.*, **20,** 488(2021).

[8] S. Yves, E. Galiffi, X. Ni, E. M. Renzi, and A. Alù. Twist-Induced Hyperbolic Shear Metasurfaces. *Phys. Rev. X.*, **14,** 021031(2024).

[9] J. L. Chen, X. Lin, M. Y. Chen, T. Low, H. S. Chen, and S. Y. Dai. A perspective of twisted photonic structures. *Appl. Phys. Lett.*, **119,** 240501(2021).

[10] L. J. Du, M. R. Molas, Z. H. Huang, G. Y. Zhang, F. Wang, and Z. P. Sun. Moire photonics and optoelectronics. *Science*, **379,** 1313(2023).

[11] S. A. Dyakov, N. S. Salakhova, A. V. Ignatov, I. M. Fradkin, V. P. Panov, J. K. Song, and N. A. Gippius. Chiral Light in Twisted Fabry-Pérot Cavities. *Advanced Optical Materials*, **12,** 2302502(2024).

[12] H. Hong, C. Huang, C. J. Ma, J. J. Qi, X. P. Shi, C. Liu, *et al.* Twist Phase Matching in Two-Dimensional Materials. *Phys. Rev. Lett.*, **131,** 233801(2023).

[13] J. Guan, J. T. Hu, Y. Wang, M. J. H. Tan, G. C. Schatz, and T. W. Odom. Far-field coupling between moire photonic lattices. *Nat. Nanotechnol.*, **18,** 514(2023).

[14] H. Y. Wang, W. Xu, Z. Y. Wei, Y. Y. Wang, Z. S. Wang, X. B. Cheng, *et al.* Twisted photonic Weyl meta-crystals and aperiodic Fermi arc scattering. *Nat. Commun.*, **15,** 2440(2024).

[15] K. C. Dong, T. C. Zhang, J. C. Li, Q. J. Wang, F. Y. Yang, Y. Rho, *et al.* Flat Bands in Magic-Angle Bilayer Photonic Crystals at Small Twists. *Phys. Rev. Lett.*, **126,** 223601(2021).

[16] B. C. Lou, N. Zhao, M. Minkov, C. Guo, M. Orenstein, and S. H. Fan. Theory for Twisted Bilayer Photonic Crystal Slabs. *Phys. Rev. Lett.*, **126,** 136101(2021).

[17] P. Wang, Y. L. Zheng, X. F. Chen, C. M. Huang, Y. V. Kartashov, L. Torner, V. V. Konotop, and F. W. Ye. Localization and delocalization of light in photonic moire lattices. *Nature*, **577,** 42(2020).

[18] H. N. Tang, F. Du, S. Carr, C. Devault, O. Mello, and E. Mazur. Modeling the optical properties of twisted bilayer photonic crystals. *Light Sci. Appl.*, **10,** 157(2021).

[19] L. Huang, W. X. Zhang, and X. D. Zhang. Moire Quasibound States in the Continuum. *Phys. Rev. Lett.*, **128,** 253901(2022).

[20] H.-Y. Luan, Y.-H. Ouyang, Z.-W. Zhao, W.-Z. Mao, and R.-M. Ma. Reconfigurable moire nanolaser arrays with phase synchronization. *Nature*, **624,** 282(2023).

[21] S. Liu, S. J. Ma, R. W. Shao, L. Zhang, T. Yan, Q. Ma, S. Zhang, and T. J. Cui. Moire metasurfaces for dynamic beamforming. *Sci. Adv.*, **8,** eabo1511(2022).

[22] H. N. Tang, B. C. Lou, F. Du, M. J. Zhang, X. Q. Ni, W. J. Xu, R. Jin, S. H. Fan, and E. Mazur. Experimental probe of twist angle-dependent band structure of on-chip optical bilayer photonic crystal. *Sci. Adv.*, **9,** eadh8498(2023).

[23] X. Y. Zhang, C. X. Bian, Z. Gong, R. X. Chen, T. Low, H. S. Chen, and X. Lin. Hybrid surface waves in twisted anisotropic heterometasurfaces. *Phys. Rev. Appl.*, **21,** 064034(2024).

[24] G. W. Hu, Q. D. Ou, G. Y. Si, Y. J. Wu, J. Wu, Z. G. Dai, *et al.* Topological polaritons and

photonic magic angles in twisted α-MoO3 bilayers. *Nature*, **582,** 209(2020).

[25] M. Z. Hasan, C. L. Kane. Colloquium: Topological insulators. *Rev. Mod. Phys.*, **82,** 3045(2010).

[26] Y. Ando, L. Fu. Topological Crystalline Insulators and Topological Superconductors: From Concepts to Materials. *Annu. Rev. Conden. Ma. P.*, **6,** 361(2015).

[27] N. P. Armitage, E. J. Mele, and A. Vishwanath. Weyl and Dirac semimetals in three-dimensional solids. *Rev. Mod. Phys.*, **90,** 015001(2018).

[28] B. H. Yan, C. Felser. Topological Materials: Weyl Semimetals. *Annu. Rev. Conden. Ma. P.*, **8,** 337(2017).

[29] L. Lu, Z. Y. Wang, D. X. Ye, L. X. Ran, L. Fu, J. D. Joannopoulos, and M. Soljacic. Experimental observation of Weyl points. *Science*, **349,** 622(2015).

[30] L. Lu, L. Fu, J. D. Joannopoulos, and M. Soljacic. Weyl points and line nodes in gyroid photonic crystals. *Nat. Photonics*, **7,** 294(2013).

[31] M. Kim, Z. Jacob, and J. Rho. Recent advances in 2D, 3D and higher-order topological photonics. *Light Sci. Appl.*, **9,** 130(2020).

[32] A. B. Khanikaev, G. Shvets. Two-dimensional topological photonics. *Nat. Photonics*, **11,** 763(2017).

[33] C. K. Chiu, J. C. Y. Teo, A. P. Schnyder, and S. Ryu. Classification of topological quantum matter with symmetries. *Rev. Mod. Phys.*, **88,** 035005(2016).

[34] C. K. Chiu, A. P. Schnyder. Classification of reflection-symmetry-protected topological semimetals and nodal superconductors. *Phys. Rev. B*, **90,** 205136(2014).

[35] S. Y. A. Yang, H. Pan, and F. Zhang. Dirac and Weyl Superconductors in Three Dimensions. *Phys. Rev. Lett.*, **113,** 046401(2014).

[36] A. Bansil, H. Lin, and T. Das. Colloquium: Topological band theory. *Rev. Mod. Phys.*, **88,** 021004(2016).

[37] T. Ozawa, H. M. Price, A. Amo, N. Goldman, M. Hafezi, L. Lu, *et al.* Topological photonics. *Rev. Mod. Phys.*, **91,** 015006(2019).

[38] S. Vaidya, J. Noh, A. Cerjan, C. Jörg, G. Von Freymann, and M. C. Rechtsman. Observation of a Charge-2 Photonic Weyl Point in the Infrared. *Phys. Rev. Lett.*, **125,** 253902(2020).

[39] H. Z. Wu, Z. Z. Liu, and J. J. Xiao. Synthetic Weyl points in plasmonic chain with simultaneous inversion and reflection symmetry breaking. *Phys. Rev. B*, **110,** 075420(2024).

[40] L. Q. Yuan, Q. Lin, M. Xiao, and S. H. Fan. Synthetic dimension in photonics. *Optica*, **5,** 1396(2018).

[41] P. Cheben, R. Halir, J. H. Schmid, H. A. Atwater, and D. R. Smith. Subwavelength integrated photonics. *Nature*, **560,** 565(2018).

[42] A. Poddubny, I. Iorsh, P. Belov, and Y. Kivshar. Hyperbolic metamaterials. *Nat. Photonics*, **7,** 948(2013).

[43] M. Kaliteevski, I. Iorsh, S. Brand, R. A. Abram, J. M. Chamberlain, A. V. Kavokin, and I. A. Shelykh. Tamm plasmon-polaritons: Possible electromagnetic states at the interface of a metal and a dielectric Bragg mirror. *Phys. Rev. B*, **76,** 165415(2007).

[44] See Supplemental Materials at [URL], which includes Refs. [45,46], for detailed derivation and simulation.

[45] A. Yariv. Photonics: Optical Electronics in Modern Communications. Oxford Univ. Press: Oxford, 2007.

[46] Q. Wang, M. Xiao, H. Liu, S. Zhu, and C. T. Chan. Measurement of the Zak phase of photonic

bands through the interface states of a metasurface/photonic crystal. Phys. Rev. B, 93, 041415(2016).

[47] C. D. Hu, Z. Y. Li, R. Tong, X. X. Wu, Z. Z. L. Xia, L. Wang, *et al.* Type-II Dirac Photons at Metasurfaces. *Phys. Rev. Lett.*, **121,** 024301(2018).

[48] C. Jörg, S. Vaidya, J. Noh, A. Cerjan, S. Augustine, G. Von Freymann, and M. C. Rechtsman. Observation of Quadratic (Charge-2) Weyl Point Splitting in Near-Infrared Photonic Crystals. *Laser & Photonics Reviews*, **16,** 2100452(2022).

[49] Q. S. Wu, A. A. Soluyanov, and T. Bzdusek. Non-Abelian band topology in noninteracting metals. *Science*, **365,** 1273(2019).

[50] Y. Yang, B. Yang, G. C. Ma, J. S. Li, S. Zhang, and C. T. Chan. Non-Abelian physics in light and sound. *Science*, **383,** 844(2024).

[51] E. C. Yang, B. Yang, O. B. You, H. C. Chan, P. Mao, Q. H. Guo, *et al.* Observation of Non-Abelian Nodal Links in Photonics. *Phys. Rev. Lett.*, **125,** 033901(2020).

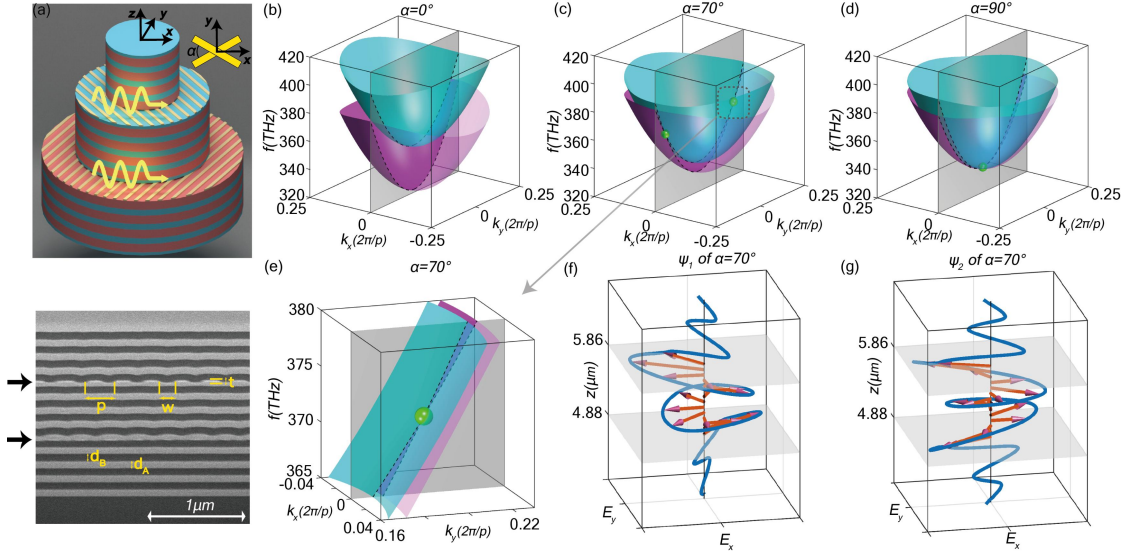

FIG. 1. (a) Schematic of twist bilayer metasurfaces embedded in 1D PCs (upper panel) and SEM image of the cross section of the structure (lower panel), the black arrows indicate the metasurfaces, $p$, $w$ and $t$ denote respectively, the period, groove and thickness of the nanostripes. (b-d) Dispersion of the TBAISs at different twist angles, the green dots stand for Dirac points. (e) Zoom in at the type-II Dirac point corresponding to the dashed region in (c). (f-g) Eigenfunctions of the TBAISs of $\alpha = 70°$ at Dirac points.

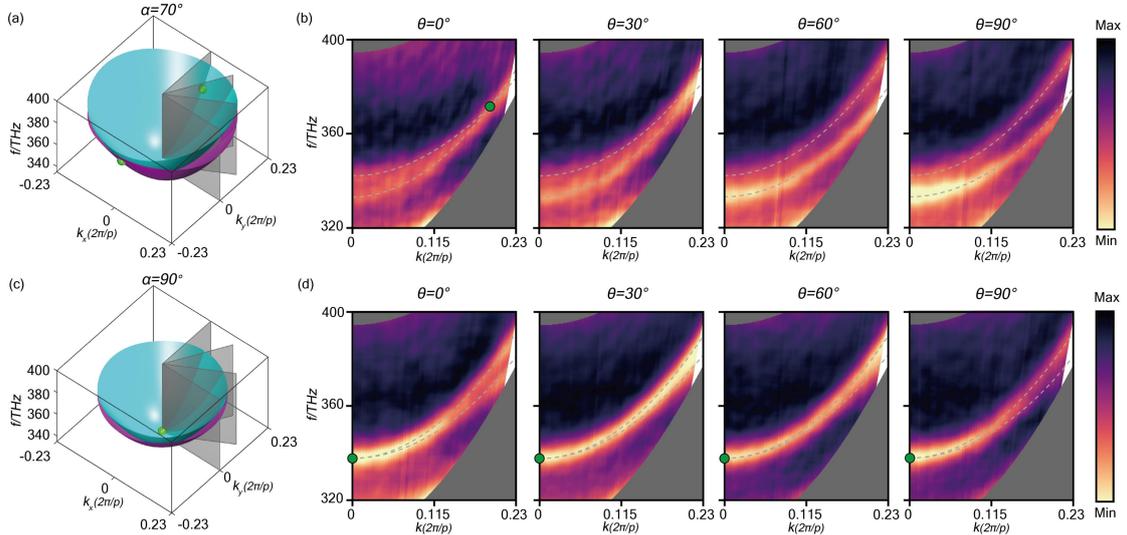

FIG. 2. (a) and (c) are the schematics of dispersion at $\alpha = 70°$ and $90°$, respectively. The degeneracies are denoted by the green dots, and the vertical gray planes indicate the azimuthal angles measured in experiment. (b) and (d) are the reflection spectra at azimuthal angles $\theta = 0°, 30°, 60°, 90°$, respectively, where the dashed lines correspond to the dispersion obtained

theoretically, and the gray shaded areas correspond to projected passband of the 1D PCs.

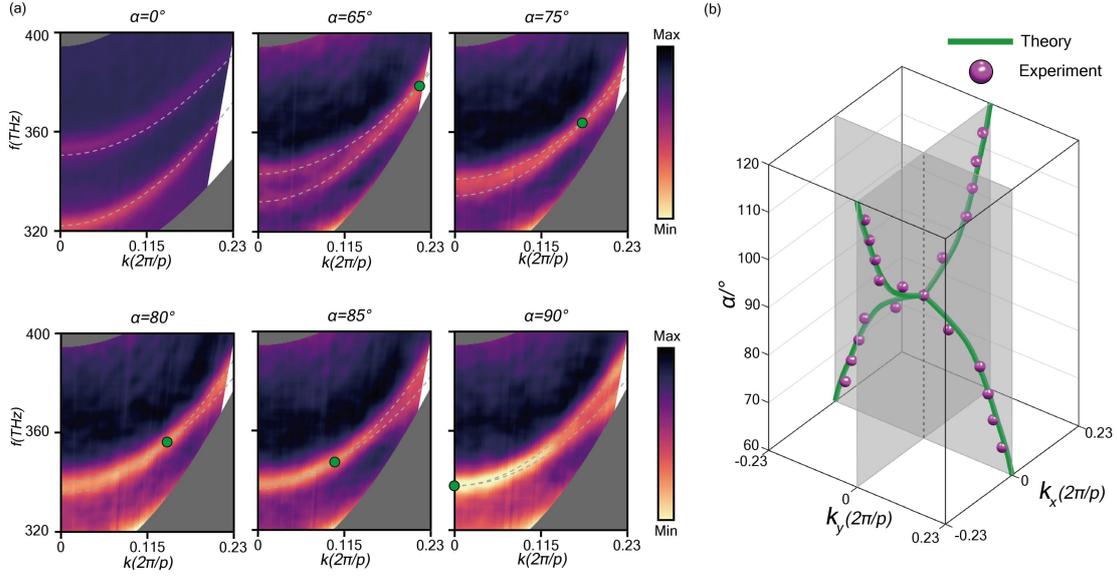

FIG. 3. (a) Reflection spectra along the $k_y$ axis at different twist angles $\alpha = 0°, 65°, 75°, 80°, 85°$ and $90°$, where the dashed lines correspond to the theoretical dispersion, and the gray shaded areas corresponds to the projected passband of the 1D PCs. (b) Theoretical nodal chain (green solid line) and corresponding experimental results (magenta solid dots).

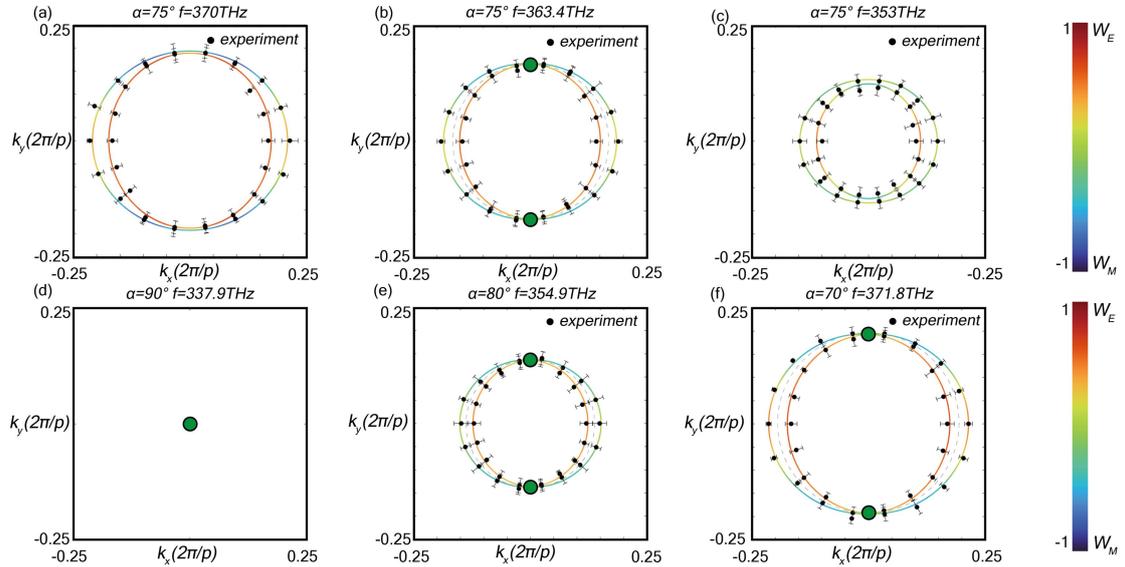

FIG. 4. (a)-(c) are the isofrequency contours above, at, below the Dirac point of twist angle $\alpha = 75°$, where the black dots represent the experimental measured dips. The colorbar is given by $(W_{E_z}^2 - W_{M_z}^2)/(W_{E_z}^2 + W_{M_z}^2)$, where $W_{E_z} = \int D_z E_z/2 \cdot dz$ and $W_{H_z} = \int B_z H_z/2 \, dz$ stand for the electric and magnetic energies along the $z$ axis. (d)-(f) are the isofrequency contours at the

Dirac points of different twist angles.

# Supplemental Material for

## Topological Degeneracy Induced by Twisting

Han Peng[1,*], Qiang Wang[1,*,†], Meng Xiao[2,3,†], Xiayi Wang[1], Shining Zhu[1] and Hui Liu[1,†]

[1]*National Laboratory of Solid State Microstructures, Jiangsu Physical Science Research Center, School of Physics, Collaborative Innovation Center of Advanced Microstructures, Nanjing University, Nanjing 210093, China*

[2]*Key Laboratory of Artificial Micro- and Nano-structures of Ministry of Education and School of Physics and Technology, Wuhan University, 430072 Wuhan, China*

[3]*Wuhan Institute of Quantum Technology, Wuhan 430206, China*

*These authors contributed equally to this work.

†Correspondence address: q.wang@nju.edu.cn; phmxiao@whu.edu.cn; liuhui@nju.edu.cn.

**S1. Using TMM to Calculate the Dispersion of TBAISs**

**S2. Properties of Single Anisotropic Interface States**

**S3. Effective Hamiltonian of TBAISs**

**S4. Derivation of Type-II and Charge 2 Dirac points**

**S5. Effects of the losses**

**S6. Detailed results of the experiment**

**S7. Experimental confirmation of Dirac points**

**S8. Details of the Full wave simulations**

## S1. Using TMM to Calculate the Dispersion of TBAISs

Transfer matrix method is adopted in our calculation of dispersions of TBAISs. The method is used to rigorously solve eigenmodes and band structure, as well as reflection and transmission spectra.

To obtain interface states, the metasurfaces here are modelled with effective permittivity tensor:

$$\begin{pmatrix} \epsilon_{xx} & 0 & 0 \\ 0 & \epsilon_{yy} & 0 \\ 0 & 0 & \epsilon_{zz} \end{pmatrix} \#(S1)$$

where $\epsilon_{xx} > 0, \epsilon_{zz} > 0, \epsilon_{yy} < 0$, and were retrieved by the S-parameter script with FDTD, Lumerical Inc. Figure. S1 depicts the effective permittivity along the $x$ and $y$ axis. Since the periodic nanostripes(grating) is pretty thin (~30nm), the specific values of the effective permittivity along the $z$ axis is almost irrelevant to the simulations (set to be unity). The imaginary part of the permittivity of the polarization that perpendicular to the stripes exhibits a peak around 580THz, corresponding to the localized resonance of the metal stripes, and is far beyond our working frequency.

Transfer matrix of isotropic media is firstly reviewed, the electric field in the isotropic media generally is:

$$\vec{E} = [(a|s^+\rangle + c|p^+\rangle)e^{ik_z z} + (b|s^-\rangle + d|p^-\rangle)e^{-ik_z z}]e^{i(k_x x + k_y y)}, \#(S2)$$

where $|s^\pm\rangle = \hat{h} = (-k_y \hat{x} + k_x \hat{y})/k_\rho$ and $|p^\pm\rangle = (\pm k_z \hat{\rho} - k_\rho \hat{z})/k$ represent two decoupled polarizations respectively, and $\hat{h}$ represents the unit vector perpendicular to the incidence plane, as $\hat{\rho}$ is the unit vector parallel to incidence plane. The superscript $\pm$ indicates forward(backward) direction. Coefficients $a, b, c, d$ of adjacent layer are connected by boundary condition.

Similarly, under principal axis, the electric field of anisotropic media is represented by:

$$\vec{E} = [a\vec{u_1}e^{iK_{1z}z} + b\vec{u_1}e^{-iK_{1z}z} + c\vec{u_2}e^{iK_{2z}z} + d\vec{u_2}e^{-iK_{2z}z}]e^{i(k_x x + k_y y)}, \#(S3)$$

where $\vec{u}_{1,2}$ and $K_{1z,2z}$ is the eigenpolarizations and its corresponding $z$ component wavevectors of each media. For a given set of $\omega, k_x, k_y$, it is proved that $K_{1z,2z}$ is solved by

$$\det\left(\begin{bmatrix} \omega^2 \mu\epsilon_x - k_y^2 - K_z^2 & k_x k_y & k_x K_z \\ k_y k_x & \omega^2 \mu\epsilon_y - k_x^2 - K_z^2 & k_y K_z \\ K_z k_x & K_z k_y & \omega^2 \mu\epsilon_z - k_x^2 - k_y^2 \end{bmatrix}\right) = 0, \#(S4)$$

and the eigenpolarization is:

$$\vec{u} = \begin{pmatrix} \dfrac{k_x}{k^2 - \omega^2 \mu\epsilon_{xx}} \\ \dfrac{k_y}{k^2 - \omega^2 \mu\epsilon_{yy}} \\ \dfrac{K_z}{k^2 - \omega^2 \mu\epsilon_{zz}} \end{pmatrix}, \#(S5)$$

where $k^2 = k_x^2 + k_y^2 + K_z^2$

Besides the transfer matrix at the boundary, the propagation matrix accounts for the phase accumulated inside each layered media is also needed, and is represented by diagonal matrix $diag(e^{iK_{1z}z}, e^{-iK_{1z}z}, e^{iK_{2z}z}, e^{-iK_{2z}z})$, for isotropic media, $K_{1z} = K_{2z} = k_z$.

The overall transfer matrix for the entire structure is calculated by multiplying the propagation matrix of each layer and the transfer matrix of each boundaries in order. After deriving the overall transfer matrix, the reflection and transmission coefficients for each polarization are available.

Now, we will focus on the eigenfrequency and eigenfunction of the multilayer structure. It is assumed that there are infinite number of unit cells in PCs above and below the structure being analyzed, transfer matrix of a unit cell is[1]:

$$M_{unit\ cell} = \begin{bmatrix} A_{TE} & B_{TE} & 0 & 0 \\ C_{TE} & D_{TE} & 0 & 0 \\ 0 & 0 & A_{TM} & B_{TM} \\ 0 & 0 & C_{TM} & D_{TM} \end{bmatrix}, \#(S6)$$

here $C_{TE} = B_{TE}^*$, $D_{TE} = A_{TE}^*$, $C_{TM} = B_{TM}^*$, $D_{TM} = A_{TM}^*$, and the eigenfunction of each polarization inside the PC is:

$$\left( \frac{B}{\frac{A+D}{2} \pm \sqrt{\left(\left(\frac{A+D}{2}\right)^2 - 1\right)} - A} \right) \#(S7)$$

Defining $M$ to be the transfer matrix of the multilayer structure considered, we have:

$$\begin{pmatrix} a_2 \\ b_2 \\ c_2 \\ d_2 \end{pmatrix} = M \begin{pmatrix} a_1 \\ b_1 \\ c_1 \\ d_1 \end{pmatrix} \#(S8)$$

Since the interface states resides in the gap of the PC, the evanescent property of modes indicates that Bloch vector at two sides of the structure is opposite, which means:

$$\frac{b_1}{a_1} = \frac{a_2}{b_2} = \frac{B_{TE}}{(A_{TE} + D_{TE})/2 \pm \sqrt{(A_{TE} + D_{TE})^2/4 - 1} - A_{TE}} = s_{TE}, \#(S9)$$

and similarly:

$$\frac{d_1}{c_1} = \frac{c_2}{d_2} = p_{TM}, \#(S10)$$

gives $\begin{pmatrix} b_2 s_{TE} \\ b_2 \\ d_2 p_{TM} \\ d_2 \end{pmatrix} = M \begin{pmatrix} a_1 \\ a_1 s_{TE} \\ c_1 \\ c_1 p_{TM} \end{pmatrix}$, after rearrangement:

$$\begin{bmatrix} M_{11} + M_{12}s_{TE} & -s_{TE} & M_{13} + M_{14}p_{TM} & 0 \\ M_{21} + M_{22}s_{TE} & -1 & M_{23} + M_{24}p_{TM} & 0 \\ M_{31} + M_{32}s_{TE} & 0 & M_{33} + M_{34}p_{TM} & -p_{TM} \\ M_{41} + M_{42}s_{TE} & 0 & M_{43} + M_{44}p_{TM} & -1 \end{bmatrix} \begin{pmatrix} a_1 \\ b_2 \\ c_1 \\ d_2 \end{pmatrix} = 0 \#(S11)$$

The matrix on the left side is defined as $D$. The equation above can only be solved if the determinant of the $D$ matrix vanishes. This matrix is dependent on the frequency $\omega$ and in-plane wavevectors $k_x$ and $k_y$. Specifically, the dispersion relation is solved implicitly by the determinant of $D(\omega, k_x, k_y)$ vanishes. In addition, the eigenfunction can be inferred from $a_1, b_2, c_1, d_2$. For demonstration, the wavefunctions at Dirac point of $\alpha = 70°$ are depicted in Fig. S2.

Now let us turn to rigorously prove there are no Dirac point at $\alpha = 0°$. If two metasurface are parallel with each other, these two interface states degenerate into symmetric and antisymmetric TE (transverse electric, and TM for transverse magnetic) modes along $k_y$ direction. Here we are going to prove the dispersion of these two interface states intersect at infinity. Consider two metasurfaces sandwiching one unitcell of PC (other conditions are proved similarly). The transfer matrix of the whole structure is:

$$M = M_{m2}M_{pc}M_{m1}, \#(S12)$$

where the $M_{pc}$ represents the transfer matrix of a unit cell of the PC, the subscript 1, 2 and $m$ stand for media 1, 2 and metasurface, and matrix $M_{mi}$ ($i = 1,2$) stands for the transfer matrix of the boundary between the metasurface and the media $i$.

The eigenmode is calculated by the eigenvectors of $A = M\sigma_x$, eigenvalues of matrix $A$ are 1 and -1, corresponding to symmetric and antisymmetric modes, the eigenfrequency is obtained by matching the eigenvectors and solutions in the band gap of the PC, and the eigenvectors in the limit of $k_y \to \infty$ are the same, meaning the eigenfrequencies of these two interface states is identical.

It is worth to point out that the single AIS is decoupled from TE-TM hybrid mode under above condition, resulting a positive overlap integral, which means a nonzero coupling strength and absence of Dirac points.

## S2. Properties of Single Anisotropic Interface States

The metasurface-PC structure at normal incidence is studied before[2], to obtain higher Q factor by inhibit radiation loss, another PC is cover on top of the metasurface-PC strucutre as shown in Fig. S3a. The top PC is composed of $b - a$ unitcells while the bottom PC is composed of $a/2 - b - a/2$, ensuring a single interface state between the bottom PC and the metasurface.

The dispersion of such structure is anisotropic owing to the nature of metasurface, 1st order term of momentum (or any odd order term) is absent due to time reversal symmetry. In addition, mirror symmetry ($x \to - x\ and\ y \to - y$) indicates the dispersion in our system is generally written as (up to 2nd term): $E = E_0 + ak_x^2 + bk_y^2$, which confirmed by transfer matrix method (as shown in Fig. S3b). Fig. S3c gives the eigenfunction of single anisotropic interface state at Γ point, which obviously are linearly polarized. Moreover, these interface states at $k_x(k_y)$ axis is TM(TE) polarized, while at other points in reciprocal space, interface states are TE-TM hybrid (see Fig. S4).

## S3. Effective Hamiltonian of TBAISs

Before proceeding, it is assumed that the electric field $E_x$ which parallel to the direction where the permittivity is negative in PC-metasurface-PC system is dominant in horizontal electric field. Therefore, the horizontal electric field is approximately aligned with the negative main axis of the metasurface. This assumption is strictly met around the $\Gamma$ point, as shown in Fig. S4, where the proportion of each horizontal polarization is depicted. Temporarily, let us assume the $x, y$ axes are the direction of the two main axis of the metasurface ($x$ and $y$ correspond to the axis where the permittivity is negative and positive), and $z$ axis is the direction where the 1D PC extends (see Fig. S4a). The approximation above dictates that the electric field on the $y$-axis (or the direction perpendicular to the negative main axis) is zero, as two polarizations cancel out on the $y$-axis. In other words, the horizontal field of the PC-metasurface-PC system is

$$E \propto \frac{k_0}{k_z}\cos(\theta)|p\rangle + \sin(\theta)|s\rangle, \#(S13)$$

where $\theta$ represent the angle between the negative main axis of the metasurface and the direction of the wavevector, $|p\rangle$ and $|s\rangle$ indicate TM and TE waves as previously defined, and the above relation holds for every layer of the PC due to the boundary condition.

Now that two metasurafaces have twisted an angle of $\alpha/2$ and $-\alpha/2$. Their eigenfunctions are

$$E_1 \propto \frac{k_0}{k_z}\cos\left(\theta - \frac{\alpha}{2}\right)|p\rangle + \sin\left(\theta - \frac{\alpha}{2}\right)|s\rangle, \#(S14)$$

$$E_2 \propto \frac{k_0}{k_z}\cos\left(\theta + \frac{\alpha}{2}\right)|p\rangle + \sin\left(\theta + \frac{\alpha}{2}\right)|s\rangle, \#(S15)$$

thus, the overlapping integral is denoted by

$$\langle E_1|E_2\rangle \propto \left(\frac{k_0}{k_z}\right)^2 \cos\left(\theta - \frac{\alpha}{2}\right)\cos\left(\theta + \frac{\alpha}{2}\right) + \sin\left(\theta - \frac{\alpha}{2}\right)\sin\left(\theta + \frac{\alpha}{2}\right)$$
$$= \cos(\alpha) + \frac{k_\rho^2}{k_z^2}\frac{1}{2}(\cos(\alpha) + \cos(2\theta)), \#(S16)$$

here $k_\rho$ is the horizontal wavevector, it is noteworthy to note that $k_z$ in above equation is a value between $k_{z1}$ and $k_{z2}$, where $k_{zi}(i = 1,2)$ indicate the vertical wavevector $k_z$ in $i$th media of the PC. According to the coupled mode method, the coupling strength is

$$J \propto \cos(\alpha) + \frac{1}{2}\frac{k_x^2 + k_y^2}{k_z^2}(\cos(\alpha) + \cos(2\theta))$$
$$= \cos(\alpha) + \frac{1}{2}\frac{k_x^2 + k_y^2}{k_z^2}\left(\cos(\alpha) + \frac{k_x^2 - k_y^2}{k_x^2 + k_y^2}\right), \#(S17)$$

the dispersion of single PC-metasurface-PC structure that rotated by angle $\alpha$ is simply represent by (note $E_0$ is omitted for convenience)

$$f(\alpha) = a[k_x\cos(\alpha) - k_y\sin(\alpha)]^2 + b[k_y\cos(\alpha) + k_x\sin(\alpha)]^2, \#(S18)$$

the overall Hamiltonian of our twisted system is:

$$H = \begin{bmatrix} f\left(-\frac{\alpha}{2}\right) & J \\ J & f\left(\frac{\alpha}{2}\right) \end{bmatrix}, \#(S19)$$

with Pauli matrices, the Hamiltonian is expressed as:

$$H = \frac{1}{2}[(a+b)(k_x^2 + k_y^2) + (a-b)(k_x^2 - k_y^2)\cos(\alpha)] \cdot \sigma_0 + (a-b)k_x k_y \sin(\alpha) \cdot \sigma_3$$
$$+ q\left[\cos(\alpha) + \frac{1}{2}[(k_x^2 + k_y^2)\cos(\alpha) + k_x^2 - k_y^2]\right] \cdot \sigma_1, \#(S20)$$

where $q$ accounts for the coupling magnitude, and reduced wavevectors are used for convenience. One might notice that the effective Hamiltonian is purely real, indicating these to resonant modes oscillate in the same phase (elaborated below).

Although the derivation of the effective Hamiltonian lacks rigor to a certain degree, it captures the essence of our twisted system, and is even in agreement with multi-metasurfaces structures. However, our model failed to predict the effect of the coupling strength $q$ since it is different for different samples. The reliability of the effective Hamiltonian can be improved by rigorously determining the polarization of the eigenfield.

The results obtained by effective Hamiltonian (Eq. 4 in the main text) are compared with rigorous results around $\alpha = \pi/2$ as shown in Fig. S5, where the parameters are shown in caption, the resemblance is satisfying.

It is pointed out above that the effective Hamiltonian is purely real in our system, indicating these to resonant modes oscillate in the same phase. Interestingly, this is true for all the mode in the gap of a 1D PC. Now we prove that all eigenmodes in the gap of 1D PC are in the same phase if the system possesses time reversal symmetry.

According to time reversal symmetry, transfer matrix is described by:

$$M = \begin{bmatrix} M_{11} & M_{12} & M_{13} & M_{14} \\ M_{12}^* & M_{11}^* & -M_{14}^* & -M_{13}^* \\ M_{31} & M_{32} & M_{33} & M_{34} \\ -M_{32}^* & -M_{31}^* & M_{34}^* & M_{44}^* \end{bmatrix}, \#(S21)$$

and

$$\begin{pmatrix} A_2 \\ B_2 \\ C_2 \\ D_2 \end{pmatrix} = \begin{pmatrix} M_{11}A_1 + M_{12}B_1 + M_{13}C_1 + M_{14}D_1 \\ M_{12}^*A_1 + M_{11}^*B_1 - M_{14}^*C_1 - M_{13}^*D_1 \\ M_{31}A_1 + M_{32}B_1 + M_{33}C_1 + M_{34}D_1 \\ -M_{32}^*A_1 - M_{31}^*B_1 + M_{34}^*C_1 + M_{33}^*D_1 \end{pmatrix}, \#(S22)$$

where the coefficient $A_i, B_i, C_i, D_i$ ($i = 1,2$) represents electric field at each side of structure considered. Such field is also the eigenfield of PC, so we have $|A_1| = |B_1|$ and $|C_1| = |D_1|$, let $B_1 = A_1 e^{i\phi_{TE}}$ and $D_1 = C_1 e^{i\phi_{TM}}$.

For TE mode:
$$A_2 e^{iz} + B_2 e^{-iz} = (M_{11}A_1 + M_{12}B_1 + M_{13}C_1 + M_{14}D_1)e^{iz}$$
$$+ (M_{12}^*A_1 + M_{11}^*B_1 - M_{14}^*C_1 - M_{13}^*D_1)e^{-iz}$$
$$= A_1 e^{i\frac{\phi_{TE}}{2}}\left[\left(M_{11}e^{-i\frac{\phi_{TE}}{2}} + M_{12}e^{i\frac{\phi_{TE}}{2}}\right)e^{iz} + c.c\right]$$
$$+ C_1 e^{i\frac{\phi_{TM}}{2}}\left[\left(M_{13}e^{-i\frac{\phi_{TM}}{2}} + M_{14}e^{i\frac{\phi_{TM}}{2}}\right) - c.c\right], \#(S23)$$

and for TM mode:

$$C_2 e^{iz} - D_2 e^{-iz} = (M_{31}A_1 + M_{32}B_1 + M_{33}C_1 + M_{34}D_1)e^{iz}$$
$$- (-M_{32}^*A_1 - M_{31}^*B_1 + M_{34}^*C_1 + M_{33}^*D_1)e^{-iz}$$
$$= A_1 e^{i\frac{\phi_{TE}}{2}}\left((M_{31}e^{-i\frac{\phi_{TE}}{2}} + M_{32}e^{i\frac{\phi_{TE}}{2}} + c.c\right)$$
$$+ C_1 e^{i\frac{\phi_{TM}}{2}}\left[\left(M_{33}e^{-i\frac{\phi_{TM}}{2}} + M_{34}e^{i\frac{\phi_{TM}}{2}}\right)e^{iz} - c.c\right], \#(S24)$$

where $c.c$ stands for complex conjugate. The electric field in layer 2 is in phase with that in layer 1 if $\arg(A_1 e^{i\frac{\phi_{TE}}{2}}) = \arg(iC_1 e^{i\frac{\phi_{TM}}{2}})$, which could be proved as follows:

Suppose our structure is embedded in infinite 1D PCs, its transfer matrix is $M$, as mention above, the coefficient at eigenfrequency satisfy:

$$\begin{pmatrix} B_2 e^{i\phi_{TE}} \\ B_2 \\ D_2 e^{i\phi_{TM}} \\ D_2 \end{pmatrix} = M \begin{pmatrix} A_1 \\ A_1 e^{i\phi_{TE}} \\ C_1 \\ C_1 e^{i\phi_{TM}} \end{pmatrix}, \#(S25)$$

for TE mode:

$$B_2 e^{i\phi_{TE}} * e^{iz} + B_2 * e^{-iz} = B_2 e^{i\frac{\phi_{TE}}{2}}\left(e^{i\frac{\phi_{TE}}{2}} e^{iz} + e^{-i\frac{\phi_{TE}}{2}} e^{-iz}\right)$$
$$= A_1 e^{i\frac{\phi_s}{2}}\left[\left(ae^{-i\frac{\phi_s}{2}} + be^{i\frac{\phi_s}{2}}\right)e^{iz} + c.c\right] + C_1 e^{i\frac{\phi_p}{2}}\left[\left(M_{13}e^{-i\frac{\phi_p}{2}} + M_{14}e^{i\frac{\phi_p}{2}}\right)e^{iz} - c.c\right], \#(S26)$$

the TE mode in the gap of the PC could always be shifted to a real number (by multiplying a phase factor). Above equation transform to:

$$B_2 e^{i\frac{\phi_{TE}}{2}}\cos\left(z + \frac{\phi_{TE}}{2}\right) = A_1 e^{i\frac{\phi_{TE}}{2}}\cos(z + \phi_1) + C_1 e^{i\frac{\phi_{TM}}{2}} i\sin(z + \phi_2), \#(S27)$$

the coefficients of these trigonometric functions are all constant, the equation holds only when:

$$\arg\left(B_2 e^{i\frac{\phi_{TE}}{2}}\right) = \arg\left(A_1 e^{i\frac{\phi_{TE}}{2}}\right) = \arg\left(iC_1 e^{i\frac{\phi_{TM}}{2}}\right) \#(S28)$$

Thus, the whole electric field in the structure is in the same phase, resulting the overlap integral purely real.

## S4. Derivation of Type-II and Charge 2 Dirac points

Firstly, we prove the existence of type-II Dirac point. By expanding the Hamiltonian around the Dirac point $k_y^{TD} = \sqrt{2\cos\alpha/(1-\cos\alpha)}, k_x = 0$, with $0° \leq \alpha \leq 90°$, keeping up to 1st order and discarding the constant term, we have

$$H = dk_y[a(1-\cos\alpha) + b(\cos\alpha + 1)]k_y^{TD} \cdot \sigma_0 + k_x(a-b)k_y^{TD}\sin\alpha \cdot \sigma_3$$
$$+ dk_y q(\cos\alpha - 1)k_y^{TD} \cdot \sigma_1, \#(S29)$$

where $dk_y = k_y - k_y^{TD}$, after performing unitary transformation $H' = VHV^{-1}$, leading to

$$H' = dk_y[a(1-\cos\alpha) + b(\cos\alpha + 1)]k_y^{TD} \cdot \sigma_0 + k_x(b-a)k_y^{TD}\sin\alpha \cdot \sigma_1$$
$$+ dk_y q(1-\cos\alpha)k_y^{TD} \cdot \sigma_2, \#(S30)$$

With

$$V = \frac{1}{\sqrt{2}}\begin{pmatrix} i & 1 \\ -i & 1 \end{pmatrix} \#(S31)$$

while the tilt parameter of the type-II Dirac point is

$$v_0 = \frac{a(1 - \cos\alpha) + b(\cos\alpha + 1)}{q(1 - \cos\alpha)}, \#(S32)$$

which is typically larger than 1 in our system.

Now let us turn to calculating the berry phase of the Charge 2 Dirac point. The effective Hamiltonian is (for convenience, the term $\sigma_0$ is omitted):

$$H = (a - b)k_x k_y \cdot \sigma_3 + q\left[\frac{1}{2}(k_x^2 - k_y^2)\right] \cdot \sigma_1, \#(S33)$$

Let $k_x = k_\rho\cos(\theta)$, $k_y = k_\rho\sin(\theta)$, one of the eigenvectors yields as

$$\frac{\phi_1}{k_\rho^2} = \left\{(b - a)\sin(2\theta) + \sqrt{q^2[\cos(2\theta)]^2 + (b - a)^2\sin(2\theta)^2}, q[-\cos(2\theta)]\right\}, \#(S34)$$

suppose $a > b$ and after normalization, one may notice that: $\phi_{\theta \to \pi/4+\epsilon} = -\phi_{\theta \to \pi/4-\epsilon}$ and $\phi_{\theta \to 5\pi/4+\epsilon} = -\phi_{\theta \to 5\pi/4-\epsilon}$, plus $\langle\phi_\theta|\partial_\theta|\phi_\theta\rangle$ is an odd function with respect to $\theta = 3\pi/4$, thus

$$\gamma = \mathrm{Im}\left(\ln\left\langle\phi_{\frac{\pi}{4}+\epsilon}\Big|\phi_{\frac{\pi}{4}-\epsilon}\right\rangle + \ln\left\langle\phi_{\frac{5\pi}{4}+\epsilon}\Big|\phi_{\frac{5\pi}{4}-\epsilon}\right\rangle\right) = 2\pi, \#(S35)$$

which means the degenerate point at Γ point when $\alpha = 90°$ is classified as charge-2 Dirac point.

## S5. Effects of the losses

The gold structure brings inevitable intrinsic loss. In addition, there can be also minor radiation loss. Our system exhibits a combined mirror symmetry $m_z m_x$ (or two-fold rotational symmetry along the y axis $C_{y2}$), and such a symmetry requires $q_{l1}(k_x, k_y) = q_{l2}(-k_x, k_y)$, where $q_{li}(k_x, k_y)$ represents the loss for the $i$-th interface state. In addition, our experiments (Fig. S6) show that $q_{li}(k_x, k_y)$ is approximately a constant near the Dirac point, i.e., $q_{li}(k_x, k_y) \cong q_{loss}$. As a result, the total Hamiltonian when considering the loss can be written as

$$H_{NH} = H_D + iq_{loss}\sigma_0, \#(S36)$$

where $H_D$ is the Hermitian Dirac Hamiltonian, and $\sigma_0$ stands for identity matrix. Thus, the Dirac point is still an observable, just shifting in the complex plane along the imaginary axis.

To experimentally evaluate the loss term (including absorption and radiation loss), we fabricated a PC-metasurface-PC structure as shown in Fig. S6. (a). Its dispersion is shown in Fig. S6. (b). We proceed to calibrate the loss near the Dirac point. Taking $\alpha = 70°$ as an example, the Dirac point occurs around $k = 0.2$, with azimuthal angle $\theta = 35°$ to the meta-grating. Thus, we measured the reflection spectrum along azimuthal angle $\theta = 33°, 35°$ and $37°$, respectively [Fig. S6(c-e)]. The measured full width at half maximum (FWHM) of each dip is 13.22THz, 13.33THz and 13.53THz as shown in Fig. S6(f)-(h). The FWHM is proportional to the loss of the interface state whose variation is pretty small ($< 2\%$ and orders of magnitude smaller than the variation of the real parts of the spectra). Correspondingly, we can regard $q_{loss}$ as approximately a constant near the Dirac point.

**S6. Detailed results of the experiment**

Alternating Ta2O5(refractive index n=2.13)/SiO2(refractive index n=1.458) layers are deposited through electron beam evaporation (AdNaNotek EBS-150U). We deposit 4 unit cells of 96nm/140nm thick Ta2O5/SiO2 on a SiO2 substrate first ($d_A = 96nm, d_B = 140nm$), then a layer of 30nm thick gold is deposit on top of the first PC. Next, using FIB system (FEI Dual Beam HELIOS NANOLAB 600i, 30keV, 40pA), a 200nm period subwavelength grating is etched on the metal layer with duty circle equals 0.5 (as the metasurface), which means the groove of the grating is $100nm$, or $p = 200nm$ and $w = 100nm$ as previously defined in the lower panel of Figure 1a. After that, 4 unit cells of 48nm/140nm/48nm thick Ta2O5/SiO2/Ta2O5 is covered on the first metasurface. The thickness of the Ta2O5 layers at both end of the middle PC (Ta2O5 layers that adjacent to the metal gratings) is set to be 55nm in order to shift the working regime to longer wavelength which in favor of our homogenous anisotropic medium model. Subsequently, another layer of metasurface is fabricated using the same technique. Lastly, the whole structure is coated by another 4 unit cells of 140nm/96nm SiO2/ Ta2O5. The samples with twist angles of 0°, 30°, 60°, 65°, 70°, 75°, 80°, 85°, 90° are examined in our experiment.

The measurement set up used in our experiment is depicted in Fig. S7. Reflection of TE incidence (electric field perpendicular to the incident plane) and TM incidence (electric field parallel to the incident plane) is obtained separately. The reflection data presented in the main text is the sum of both polarizations, and the error bars are obtained by the half width of the corresponding dips in Figure 4. The permittivity of metasurface with larger incident angle are set smaller than normal incidence due to the finite period of the metasurface. Lastly, due to fabrication errors, all the theoretical spectra are redshifted by 17THz to match with the experimental results.

Figure. S8a shows the reflection spectra at normal incidence for various twist angle, the incident polarization is parallel to the diagonal line of two metasurface ($x$ axis in the inset of figure. 1a), the experiment results are in accord with theoretical predictions as shown in Fig. S8b. Figures S9 and S10 show the theoretical and experimental results of various scan direction at $\alpha = 70°$ and $k_y$ axis at various twist angle.

For direct observation of isofrequency contour, Fourier transform is used (Figure S7b), isofrequency contours for fixed frequency at one polarization are observed at fourier space and are depicted in Fig. S11, accompanied by theoretical predicted contours plotted in grey dashed lines.

**S7. Experimental confirmation of Dirac points**

It is known from numerical results that the two bands near the Dirac points correspond to different polarizations, therefore, when determining the Dirac points, the positions of closest two dips corresponding to two polarizations in frequency domain are regarded as the degenerate point, as shown in figure S12, where $\alpha = 75°$ is given as an example in the upper two rows, the reflection spectrum at specific wavevectors is shown in figure S12. (d)-(f), corresponding to the three vertical lines in figure S12. (b) and (c), it is obvious that two dips are separated at $k = 0.124$ and $k = 0.212$, and these two dips met in frequency at $k = 0.168$, where regarded as the degenerate point, similar method is adopted for different twist angle in Figure S12. (g)-(i).

However, the experimental confirmation of TDs at smaller angle is limited by the experimental setup. Specifically, the numerical aperture of the objective lens in the Angular Resolved Microscopic Spectroscopy (ARMS, Ideaoptics Inc.) is approximately $0.8660$, which corresponds to a maximum incident angle at about 60°. The corresponding in-plane wavevector $k$ can be measured at 460THz is $0.25 * 2\pi/p$. In comparison, for the sample with $\alpha = 30°$, the Dirac points are at $k = 0.34 * 2\pi/p$ with effective Hamiltonian (shown as green dot in fig. S13), which is far above the maximum in-plane wavevector that we can measured. Thus, the TDs with smaller twist angle ($\alpha < 60°$) cannot be identified with our current experiment setup.

In addition, we employed a direct simulation with *COMSOL Multiphysics*, there was no significant difference between the results from the *COMSOL* and from the effective media approximation for twist angle as low as $\alpha = 30°$. However, for near zero twist angle, the supercell of two subwavelength grating may jeopardize the effective medium approximation. It should be pointed out that the main concern here is the twist angle near 90°, where the approximation is reasonably valid.

**S8. Full wave simulations of the Birefringence effects**

The direct experimental observation of birefringence effects is inhibited by the limited coupling between the TBATSs and the plane wave in free space, therefore, it is difficult to detect the interface states outside the structure using current technique. To emulate the birefringence effect, full-wave simulations of the birefringence effect are performed with *COMSOL Multiphysics* v.5.3. We choose the PC-metal-PC structure as an isotropic media, with the permittivity of the metal set to $-9$, the thicknesses of each layers are matched to the twisted PC-metasurface-PC-metasurface-PC structure in the main text, and the loss terms in the metasurface are set to zero for demonstration purpose. The isotropic structure hosts isotropic interface states, which are used as 2D light beam to inject into the twist structure. Specifically, the interface mode at frequency 372THz with an effective refractive index $n_{eff} = 0.480442$ is injected into the $y$ axis of the twisted structure with $\alpha = 80°$. A serie of plane waves are simulated and combined as an effective gaussian beam. The wavevector perpendicular to the gaussian beam (or transverse wavevector) sweeps from -0.02 to 0.02 (in the unit of $2\pi/200nm$), while the amplitude of each plane wave is $w_0 \exp\left(-k_t^2 w_0^2\right)/\sqrt{\pi}$, where $w_0$ and $k_t$ represent the half width of gaussian beam and the transverse wavevector. To diminish radiation loss, the number of unitcells of outer two PCs is set to be 15. The amplitudes of the refracted waves are then collected at certain z positions. Figure S14. gives the full wave simulation of normal incidence and 45° incidence on the twisted structure, the latter clearly shows two refracted waves.

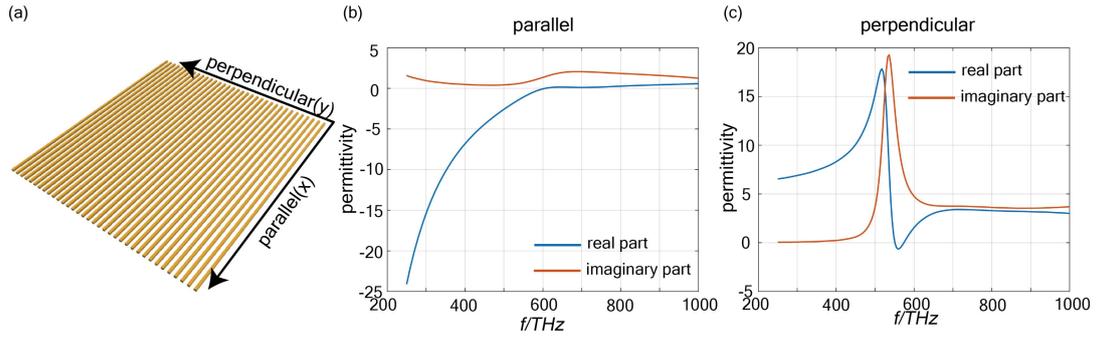

FIG. S1. (a). Schematic of subwavelength periodic nanostripes. (b) and (c) are the effective permittivity along the parallel and perpendicular direction.

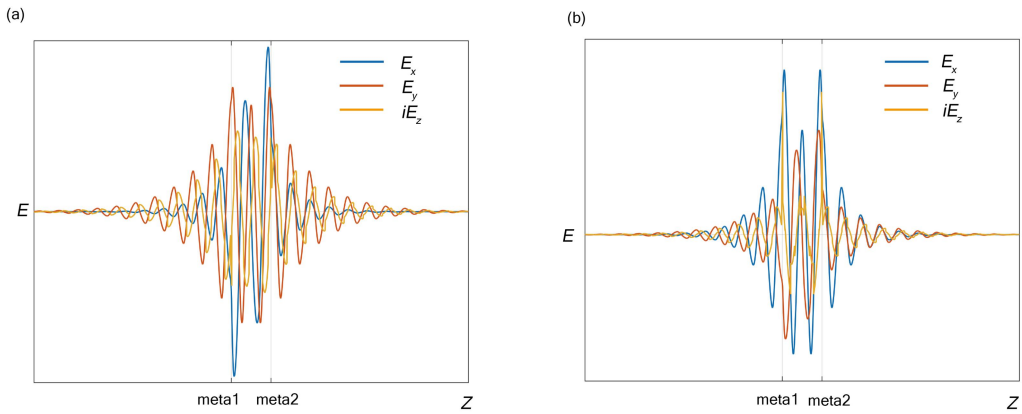

FIG. S2. (a) and (b) are the wavefunctions of two eigenstates at Dirac point of $\alpha = 70°$

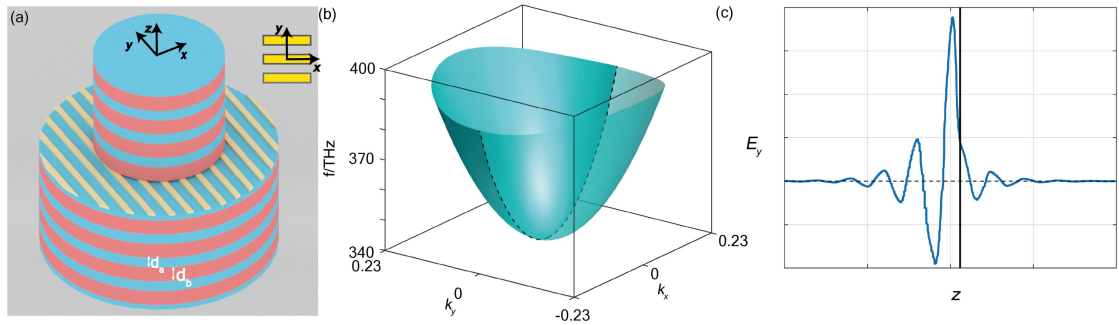

FIG. S3. (a). Schematic of PC-metasurface-PC structure. (b). Dispersion of the AIS in a PC-metasurface-PC structure. (c) Eigenfunction of AIS, here the solid black line indicates the position of the metasurface.

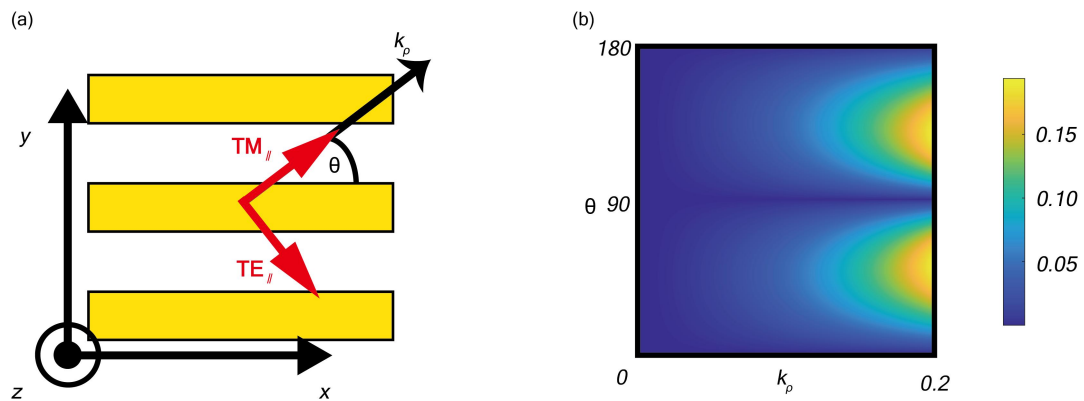

FIG. S4. (a). Schematic of horizontal field of two polarizations of specific direction of $k_\rho$. (b). $\int |E_y|dz / \int |E_x|dz$ at different $k_\rho$ of various $\theta$

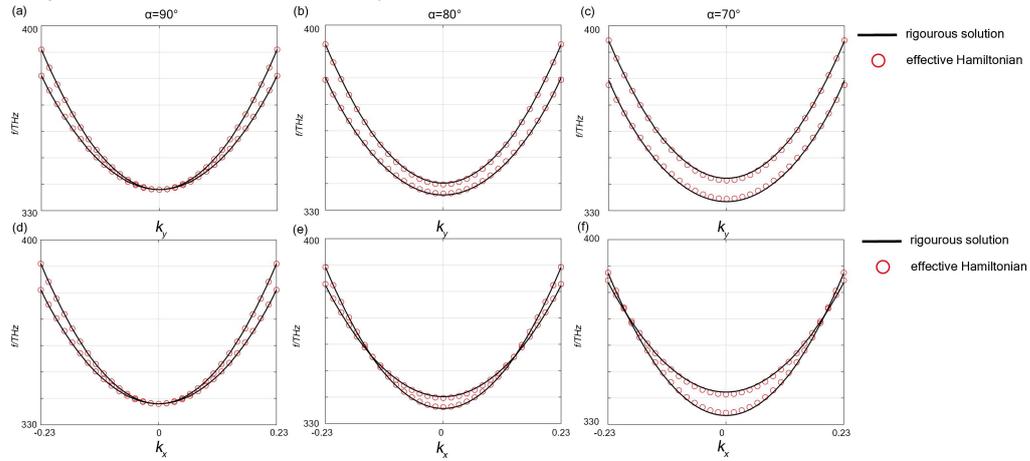

FIG. S5. Comparison between rigorous solution and the effective Hamiltonian. The coefficients are $f_0 = 337.9$, $q = 10.027$, $a = 51.78$, $b = 45.55$, the reduced constant of wavevector is 4.3234.

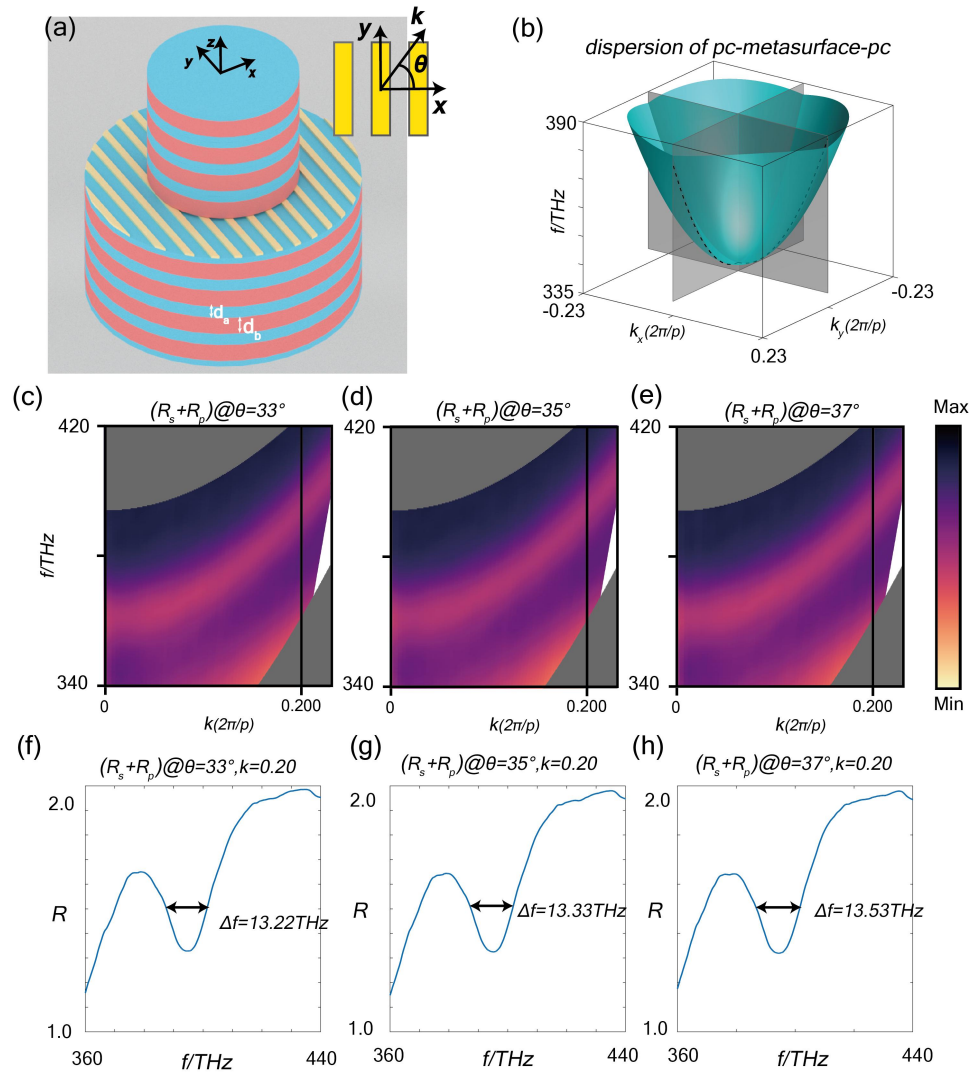

FIG. S6. (a). Schematic of a pc-metasurface-pc composite system, where the inset shows the definition of azimuthal angle $\theta$. The parameters used are same as Fig. 1(a) in the main text except that here only one metasurface is fabricated. (b). Dispersion of the anisotropic interface state. (c)-(e). Measured reflection spectra along $\theta = 33°, 35°$ and $37°$. (f)-(h). Reflection spectrum at $k = 0.2$ as indicated by the black solid lines in (c)-(e), the loss term corresponds to half width of the dip (i.e., $\Delta f/2$ ).

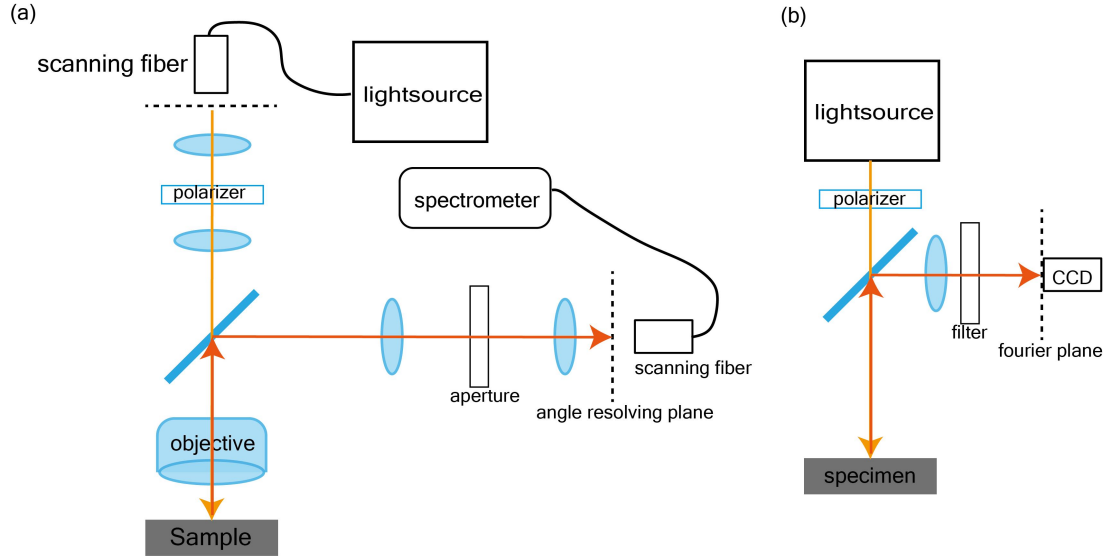

FIG. S7. (a) Experimental set up for angular resolved microscope spectroscopy (ARMS, Ideaoptics Inc.). (b) Experimental set up for the observation of isofrequency contours.

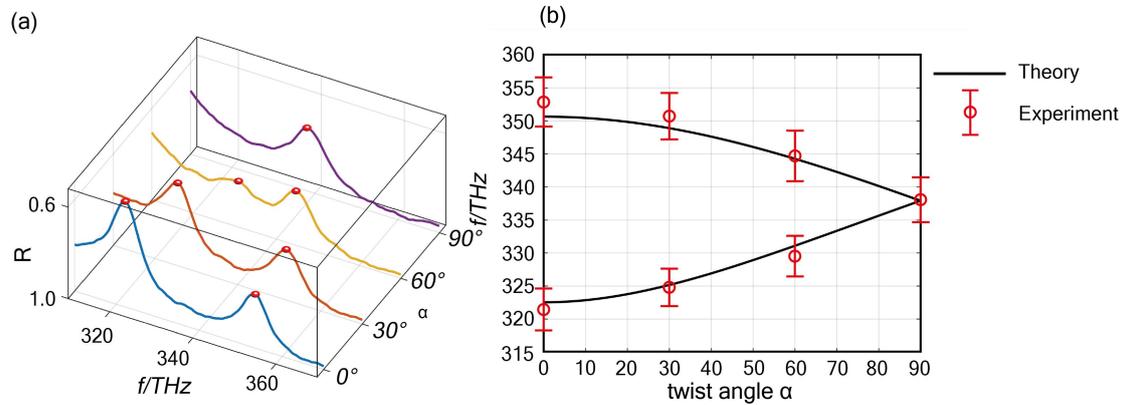

FIG. S8. (a) Experimental reflection spectra of normal incidence at different twist angle $\alpha$. (b). Comparison between theoretical result (black solid line) and experimental result (red circles) in (a), the error bars indicate the half width of corresponding dips.

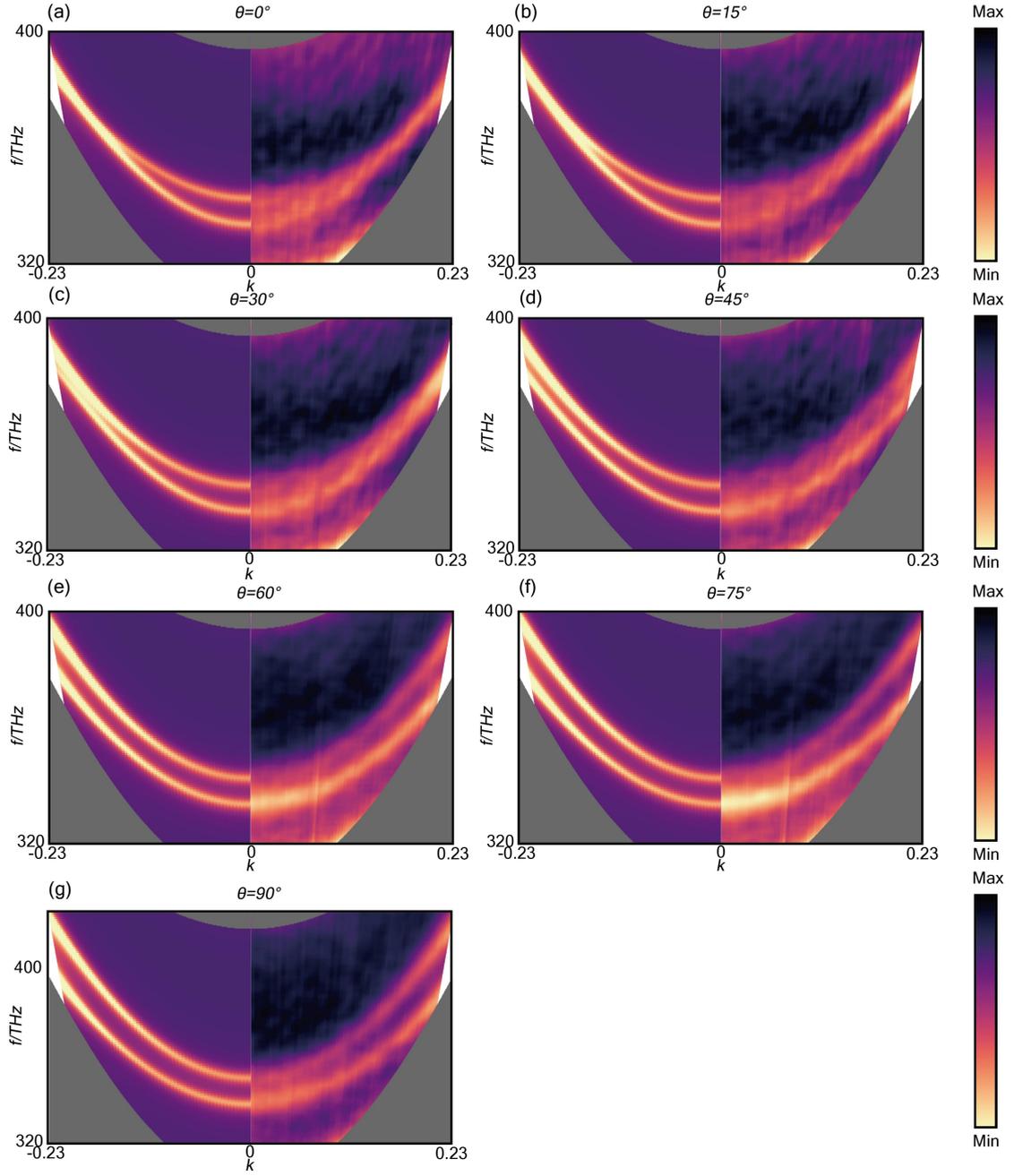

FIG. S9. Theory (left panel) vs. experiment (right panel) of twist angle $\alpha = 70°$. (a)-(g) represent the reflection spectra of azimuthal angle $\theta = 0°, 15°, 30°, 45°, 60°, 75°, 90°$ respectively.

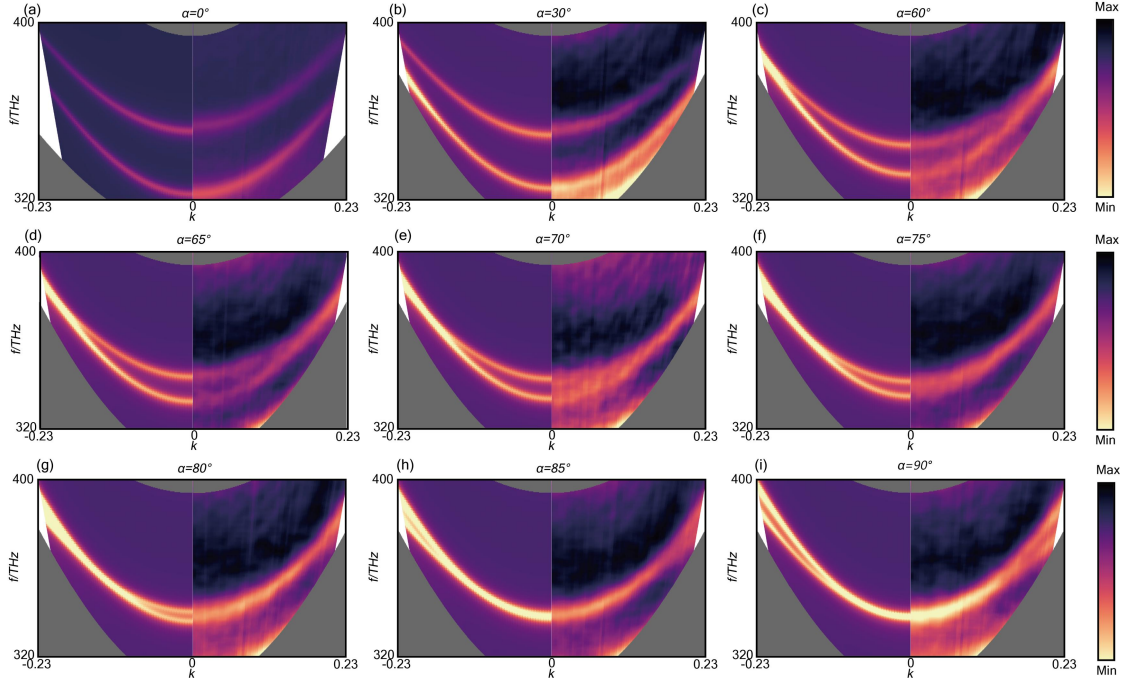

FIG. S10. Theory (left panel) vs. experiment (right panel) along the $k_y$ axis. (a)-(i) represent the reflection spectra of $\alpha = 0°$, $30°$, $60°$, $65°$, $70°$, $75°$, $80°$, $85°$, $90°$ respectively.

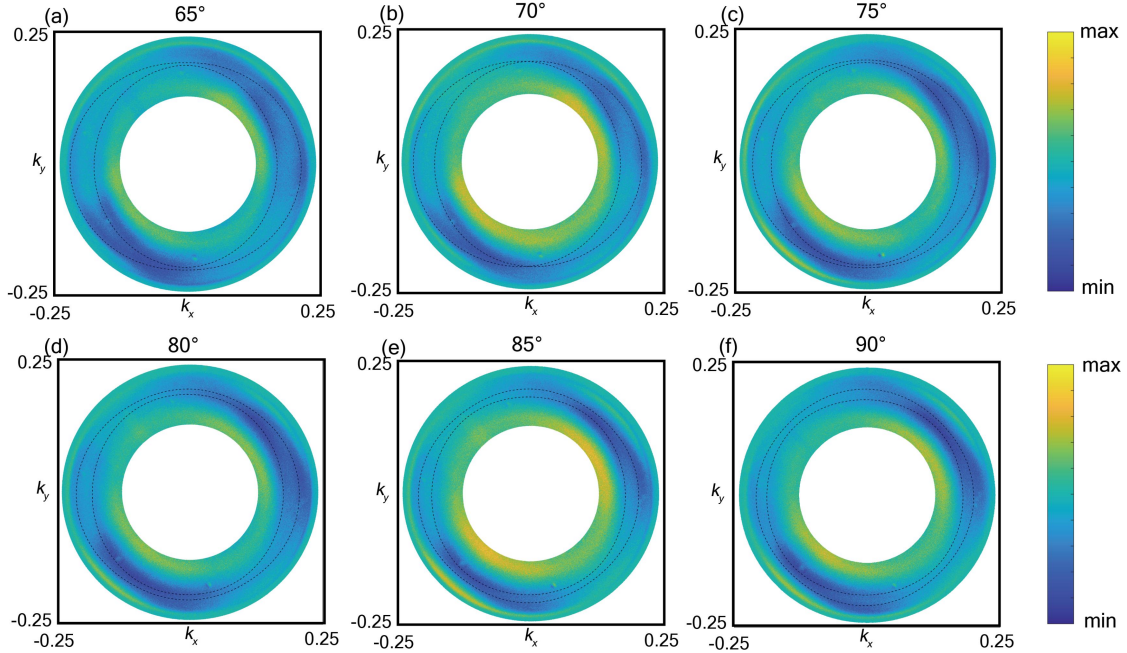

FIG. S11. (a)-(f) Experimental observation of isofrequency contours of different twist angle at 810 nm, the center regions are cut out for better comparison, and the black dashed lines indicate the isofrequency contours in theory.

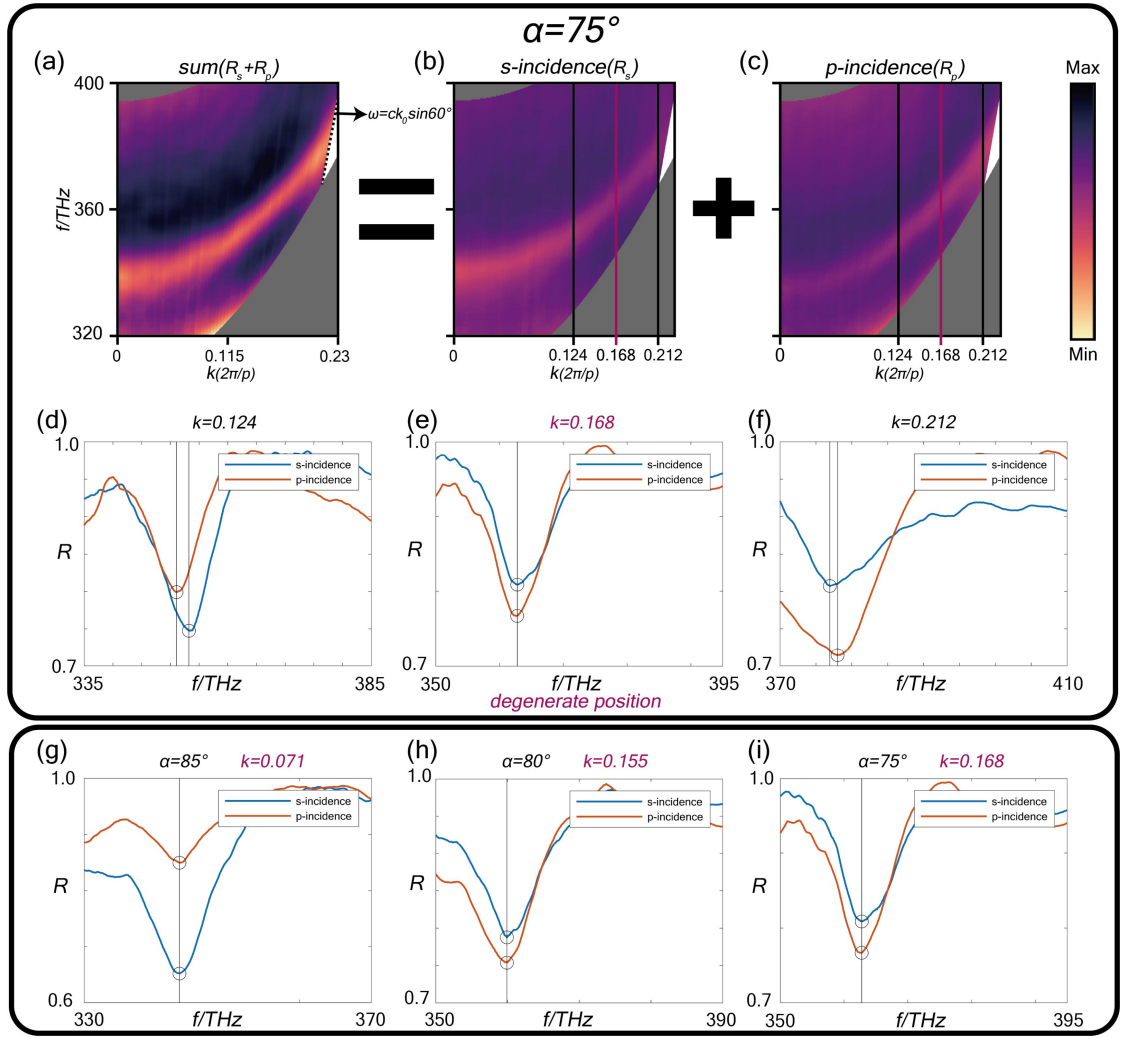

FIG. S12. (a)-(c) experimental reflection spectrum of different polarization of $\alpha = 75°$. (d)-(e) reflection spectrum at different wavevector of $\alpha = 75°$. (g)-(i) reflection spectrum at degenerate positions of various twist angles.

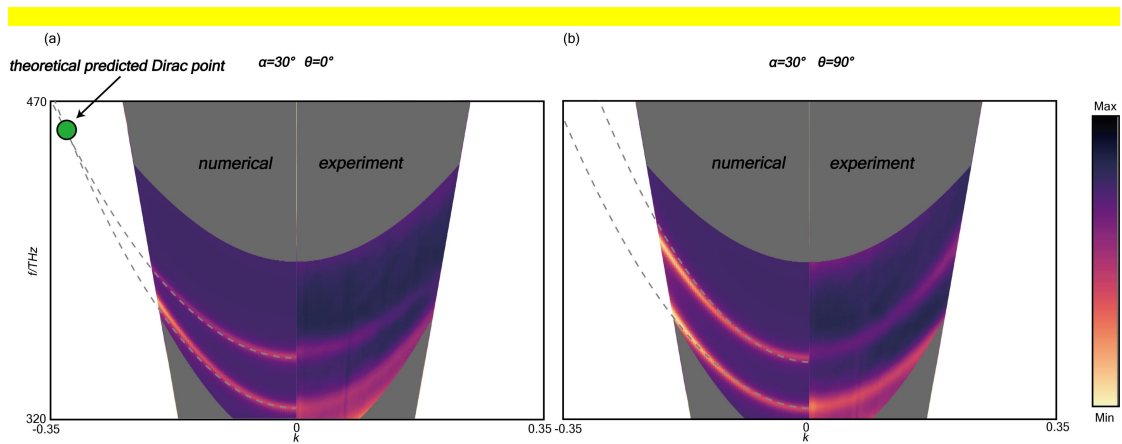

FIG. S13. Numerical (right panel) vs. experiment (left panel) with a twist angle $\alpha = 30°$, where the green dot represents the theoretical predicted Dirac point, which is far beyond our experimental scope. Here same as before, $\theta$ is the azimuthal angle of the incident plane in (a) and (b).

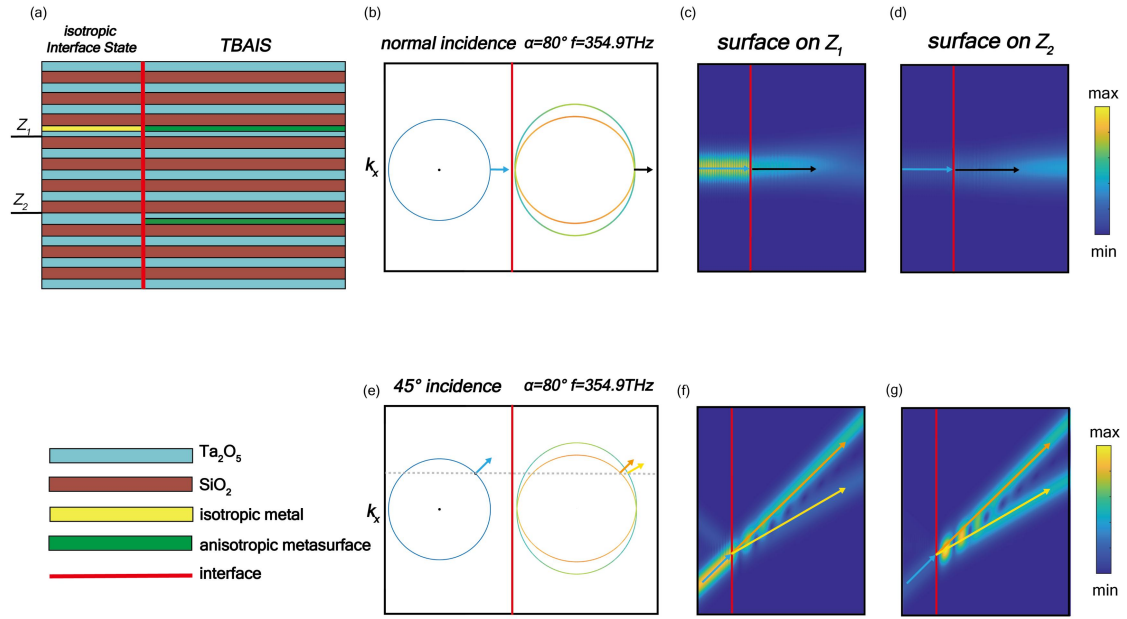

FIG. S14. Birefringence effects of TBAIS. (a) Configuration of incidence on TBAIS. (b) and (e) are the isofrequency contours of isotropic interface states and TBAIS, the arrows indicate directions of incident and refracted waves. (c) and (d) are the full-wave simulations of normal incidence for two different $z$ planes. (f) and (g) are the full-wave simulations of 45° incidence for two different $z$ planes.

**reference**

[1] A. Yariv. *Photonics: Optical Electronics in Modern Communications*. Oxford Univ. Press: Oxford, 2007.

[2] Q. Wang, M. Xiao, H. Liu, S. Zhu, and C. T. Chan. Measurement of the Zak phase of photonic bands through the interface states of a metasurface/photonic crystal. *Phys. Rev. B*, **93,** 041415(2016).